\documentclass[11pt,a4paper]{article}
\usepackage[english]{babel}
\usepackage{amsmath,amsthm,amssymb,epsfig,latexsym}
\usepackage{color}
\usepackage{ulem}
\usepackage[numbers,square,sort&compress]{natbib}

\newcommand{\la}{u}
\newcommand{\muu}{v}

\newcommand{\as}{\lambda}
\newcommand{\bla}{\bar u}
\newcommand{\bmu}{\bar v}

\newcommand{\rr}{r_0}

\def\Izer{{\sf K}}
\newcommand{\be}[1]{\begin{equation}\label{#1}}
\newcommand{\ba}[1]{\begin{multline}\label{#1}}
\newcommand{\ee}{\end{equation}}
\newcommand{\ea}{\end{eqnarray}}

\newcommand{\diag}{\mathop{\rm diag}}

\newcommand{\tr}{\mathop{\rm tr}}

\newcommand{\Sym}{\mathop{\rm Sym}\limits}
\def\Izer{{\sf K}}
\newcommand{\cc}{\varkappa}

 \makeatletter
 \@addtoreset{equation}{section}
 \makeatother
 
\newcommand{\bea}{\begin{eqnarray}}
\newcommand{\eea}{\end{eqnarray}}

\begin{document}

\begin{center}
\begin{LARGE}
{\bf One-dimensional two-component Bose gas\\and the algebraic Bethe ansatz }
\end{LARGE}

\vspace{40pt}

\begin{large}
{N.~A.~Slavnov\footnote{nslavnov@mi.ras.ru}}
\end{large}

 \vspace{12mm}

\vspace{4mm}

{\it Steklov Mathematical Institute,
Moscow, Russia}

\end{center}


\vspace{4mm}


\begin{abstract}
We apply the nested algebraic Bethe ansatz to a model of one-dimensional two-compo\-nent Bose gas with $\delta$-function repulsive interaction.
Using a lattice approximation of the $L$-operator we find Bethe vectors of the model in the continuous limit. We also obtain a series
representation for the monodromy matrix of the model in terms of Bose fields. This representation allows us to study an asymptotic
expansion of the monodromy matrix over the spectral parameter.
\end{abstract}

\vspace{1cm}

\vspace{2mm}

\section{Introduction}

In this paper we  consider a model of one-dimensional two-component Bose gas with $\delta$-function repulsive interaction (TCBG model).
This model is a generalization of the Lieb--Liniger model \cite{LiebL63,Lieb63} (Quantum nonlinear Schr\"odinger equation), in which Bose fields
have two internal degrees of freedom (colors). This model was solved by C.~N.~Yang \cite{Yang67} where
the eigenvectors and the spectrum of the Hamiltonian were found.  The general approach to the solution of the
model with $n$ internal degrees of freedom (multi-component Bose gas) was given in \cite{Sath68} (see also
\cite{Sath75,Gaud83}). 
The nested algebraic Bethe ansatz was applied to this model in \cite{Kul81,KulRes82}.
The main goal of this paper is to create a base for  calculating from factors of local operators in this model in the framework of the
nested algebraic Bethe ansatz.

The algebraic Bethe ansatz is an efficient  method for finding the spectra of quantum Hamiltonians.
However, in a viewpoint of calculating form factors of local operators application of this method meets some difficulties.
The main problem is to embed the local operators of the
model under consideration into the algebra of  the monodromy matrix entries. In some cases, this problem can be solved \cite{KitMT99,MaiT00}.
However, to construct such a solution requires that the monodromy matrix of the model $T(u)$ would be expressed in terms of the $R$-matrix. This is not the case of the TCBG model. On the other hand, in the framework of the traditional approach  one can easily obtain representations for form factors of local operators and correlation functions in terms of multiple integrals of the product of the wave functions. However, the evaluation of those multiple integrals is facing serious technical difficulties, and still they have been computed only for some relatively simple special cases \cite{PozOK12}.

Recently  a method of calculating form factors of local operators in models possessing $GL(3)$ symmetry  was developed
in \cite{PakRS15c}. This method is based on the nested algebraic Bethe ansatz and deals with partial zero modes of the monodromy matrix entries $T_{ij}(u)$
\cite{PakRS15a} in a composite model \cite{IzeK84}. Most of the tools of this approach can be directly used in TCBG model, however some of them
should be slightly modified. In particular, one should adjust a definition of the zero modes. We solve these problems in the present paper.

We consider a lattice approximation of the TCBG model. Using the $L$-operator obtained in \cite{Kul81,KulRes82} we construct a monodromy
matrix and Bethe vectors. We show that  these vectors have a correct  continuous limit. We also obtain an explicit series representation for the
monodromy matrix in terms of local Bose fields. Using this representation we are able to derive an asymptotic expansion of the monodromy matrix over the spectral parameter. In this way we find the zero modes.

The paper is organized as follows.  In section~\ref{S-ABA} we describe a general scheme of the algebraic Bethe ansatz. We define
Bethe vectors of $GL(3)$-invariant models and give their representation in a multi-composite model. Section~\ref{S-TC-BG}  is devoted to a brief
description the TCBG model. In section~\ref{S-LTC-BG} we give a lattice approximation of the TCBG model in the framework of the nested algebraic Bethe ansatz.    In section~\ref{S-BV-LO}  we consider continuous limit of the Bethe vectors of the lattice model. In section~\ref{S-RMM-BF} we obtain a series
representation of the TCBG monodromy matrix. Using this representation we find an antimorphism between Bose fields in section~\ref{S-MF} and zero modes
of the monodromy matrix entries in section~\ref{S-ZM}. In conclusion we discuss some further applications of the results obtained.

\section{Algebraic Bethe ansatz\label{S-ABA}}

In this section we describe an abstract scheme of the algebraic Bethe ansatz, which is valid  for a wide
class of quantum integrable models \cite{FadST79,FadT79,FadLH96}.
The key objects of the algebraic Bethe ansatz are a monodromy matrix and $R$-matrix.
The models considered below are described by the $GL(3)$-invariant
$R$-matrix \cite{KulR81,KulR83} acting in the tensor product $V_1\otimes V_2$ of two auxiliary spaces
$V_k\sim\mathbb{C}^3$, $k=1,2$:
 \be{R-mat}
 R(x,y)=\mathbf{I}+g(x,y)\mathbf{P},\qquad g(x,y)=\frac{c}{x-y}.
 \ee
In the above definition, $\mathbf{I}$ is the identity matrix in $V_1\otimes V_2$, $\mathbf{P}$ is the permutation matrix
that exchanges $V_1$ and $V_2$, and $c$ is a constant.

The monodromy matrix $T(w)$ satisfies the algebra
\be{RTT}
R_{12}(w_1,w_2)T_1(w_1)T_2(w_2)=T_2(w_2)T_1(w_1)R_{12}(w_1,w_2).
\ee
Equation \eqref{RTT} holds in the tensor product $V_1\otimes V_2\otimes\mathcal{H}$,
where $\mathcal{H}$ is the Hilbert space of the Hamiltonian of the model under consideration.
The  matrices $T_k(w)$ act non-trivially in
$V_k\otimes \mathcal{H}$.  We assume that the  space $\mathcal{H}$ possesses a
 pseudovacuum vector $|0\rangle$. Similarly the dual space $\mathcal{H}^*$ possesses
 a dual pseudovacuum vector $\langle0|$. These vectors
are annihilated by the operators $T_{ij}(w)$, where $i>j$ for  $|0\rangle$ and $i<j$ for $\langle0|$.
At the same time both vectors are eigenvectors of the diagonal entries of the monodromy matrix
 \be{Tjj}
 T_{ii}(w)|0\rangle=\lambda_i(w)|0\rangle, \qquad   \langle0|T_{ii}(w)=\lambda_i(w)\langle0|,\quad i=1,2,3,
 \ee
where $\as_i(w)$ are some scalar functions. In the framework of the general scheme of the algebraic Bethe ansatz $\as_i(w)$ remain free functional parameters. Actually, it is always possible to normalize
the monodromy matrix $T(w)\to \as_2^{-1}(w)T(w)$ so as to deal only with the ratios
 \be{ratios}
 r_1(w)=\frac{\lambda_1(w)}{\lambda_2(w)}, \qquad  r_3(w)=\frac{\lambda_3(w)}{\lambda_2(w)}.
 \ee
Below we assume that $\lambda_2(w)=1$.

The trace in the auxiliary space $V\sim\mathbb{C}^3$ of the monodromy matrix $\tr T(w)$ is called the transfer matrix. It is a generating
functional of the Hamiltonian and all integrals of motion of the model.

\subsection{Notation}
We use the same notation and conventions as in the papers \cite{BelPRS13a,PakRS14b}.
Besides the function $g(x,y)$ we also introduce a function $f(x,y)$
\be{univ-not}
 f(x,y)=1+g(x,y)=\frac{x-y+c}{x-y}.
\ee

We denote sets of variables by bar: $\bar w$, $\bla$, $\bmu$ etc.
Individual elements of the sets are denoted by subscripts: $w_j$, $\la_k$ etc.  Notation $\bla_i$,
means $\bla\setminus u_i$ etc. We also consider partitions of
sets into disjoint subsets and denote them by symbol $\Rightarrow$. Subsets are denoted by superscripts
in parenthesis: $\bla^{(j)}$.
For example, the notation $\bar u\Rightarrow\{\bar u^{(1)},\;\bar u^{(2)}\}$ means that the
set $\bar u$ is divided into two disjoint subsets $\bar u^{(1)}$ and $\bar u^{(2)}$, such that
$\bar u^{(1)}\cap\bar u^{(2)}=\emptyset$ and $\{\bar u^{(1)},\bar u^{(2)}\}=\bar u$.

In order to avoid too cumbersome formulas we use a shorthand notation for products of  operators or
functions depending on one or two variables. Namely, if the operators $T_{ij}$ or the functions $r_k$ \eqref{ratios}  depend
on sets of variables, this means that one should take the product over the corresponding set.
For example,
 \be{SH-prod}
 T_{ij}(\bla)=\prod_{\la_k\in\bla} T_{ij}(\la_k);\quad
 r_3(\bla^{(1)})=\prod_{u_j\in\bla^{(1)}} r_3(u_j).
 \ee
Similar convention is applied to the products of the functions $f(x,y)$:
 \be{SH-prod1}
  f(z, \bar w_i)= \prod_{\substack{w_j\in\bar w\\w_j\ne w_i}} f(z, w_j);\quad
 f(\bla,\bmu)=\prod_{u_j\in\bla}\prod_{v_k\in\bmu} f(u_j,v_k).
 \ee

\subsection{Bethe vectors\label{SS-BV}}

The eigenvectors of the transfer matrix are
called on-shell Bethe vectors (or simply on-shell vectors). In order to find them one should first construct
generic Bethe vectors. In  the framework of the algebraic Bethe ansatz generic Bethe vectors are polynomials
in operators $T_{ij}$ with $i<j$ applied to the pseudovacuum vector. We denote them by $\mathbb{B}_{a,b}(\bla;\bmu)$, stressing
that they are  parameterized by two sets of
complex parameters $\bla=\{\la_1,\dots,\la_a\}$ and $\bmu=\{\muu_1,\dots,\muu_b\}$ with $a,b=0,1,\dots$.
Different representations for Bethe vectors  were found in \cite{TarVar93,KhoPakT07,KhoPak08,BelPRS12c}.
We give here one of the representations obtained in \cite{BelPRS12c}
\be{BV-expl}
\mathbb{B}_{a,b}(\bla;\bmu) =\sum \frac{\Izer_{n}(\bmu^{(1)}|\bla^{(1)})}{f(\bmu,\bla)}
f(\bmu^{(2)},\bmu^{(1)})f(\bla^{(1)},\bla^{(2)})\,
T_{13}(\bmu^{(1)})T_{23}(\bmu^{(2)})T_{12}(\bla^{(2)})|0\rangle.
\ee
Here the sums are taken over partitions of the sets $\bar u\Rightarrow\{\bar u^{(1)},\bar u^{(2)}\}$ and $\bar v\Rightarrow\{\bar v^{(1)},\bar v^{(2)}\}$ with $0\leq\#\bar u^{(1)}=\#\bar v^{(1)}=n\leq\mbox{min}(a,b)$. We recall that the notation $T_{13}(\bla^{(1)})$ (and similar ones) means the product of the
operators $T_{13}(u)$ with respect to the subset $\bla^{(1)}$.
Finally, $ \Izer_n(\bar v^{(1)}|\bar u^{(1)})$ is the the partition function of the six-vertex model with domain wall boundary conditions  \cite{Kor82}. Its explicit representation was found in \cite{Ize87}
\begin{equation}\label{K-def}
\Izer_n(\bar x|\bar y)=\left(\prod_{1\le k<j\le n}g(x_j,x_k)g(y_k,y_j)\right)
\frac{f(\bar x,\bar y)}{g(\bar x,\bar y)}
\det_n \left(\frac{g^2(x_j,y_k)}{f(x_j,y_k)}\right).
\end{equation}
In particular, $\Izer_1(x|y)=g(x,y)$.

A generic Bethe vector becomes on-shell, if the  parameters $\bla$ and $\bmu$
satisfy a system of Bethe equations:
\be{AEigenS-1}
\begin{aligned}
r_1(u_{i})&=\frac{f(u_{i},\bla_{i})}{f(\bla_{i},u_{i})}f(\bmu,u_{i}),\qquad i=1,\dots,a,\\
r_3(v_{j})&=\frac{f(\bmu_{j},v_{j})}{f(v_{j},\bmu_{j})}f(v_{j},\bla), \qquad j=1,\dots,b.
\end{aligned}
\ee
Recall that $\bla_{i}=\bla\setminus u_i$ and $\bmu_j=\bmu\setminus v_j$.

\subsection{Multi-composite model}

Study of the properties of local operators in the framework of the algebraic Bethe ansatz can be done by the  use of a composite model \cite{IzeK84}. Suppose that we have a lattice quantum model of $N$ sites. Then the monodromy matrix $T(u)$ is a product
of local $L$-operators
\be{Model}
T(u)=L_N(u)\dots L_1(u).
\ee
Let us fix an arbitrary site $m$, $1\le m\le N$. Then \eqref{Model} can be written as
\be{T-TT}
T(u)=T^{(2)}(u)T^{(1)}(u),
\ee
where
\be{Comp-Model}
T^{(1)}(u)=L_m(u)\dots L_1(u), \qquad T^{(2)}(u)=L_N(u)\dots L_{m+1}(u).
\ee

Representation \eqref{T-TT} defines a composite model. In the framework of the composite model the original matrix $T(u)$ is called the total
monodromy matrix, while the matrices $T^{(2)}(u)$ and $T^{(1)}(u)$ are called partial monodromy matrices. The matrix elements of the partial monodromy matrices $T^{(1)}(u)$ and $T^{(2)}(u)$  act in the spaces $\mathcal{H}^{(1)}$ and $\mathcal{H}^{(2)}$  associated
to the lattice intervals $[1,m]$ and $[m+1,N]$ respectively. The entries of the total monodromy matrix act in the  space of states
$\mathcal{H}=\mathcal{H}^{(1)}\otimes \mathcal{H}^{(2)}$.

In the framework of the algebraic Bethe ansatz it is assumed that $\mathcal{H}^{(1)}$ and $\mathcal{H}^{(2)}$ possess pseudovacuum vectors
$|0\rangle^{(k)}$, $k=1,2$, such that
$|0\rangle=|0\rangle^{(1)}\otimes |0\rangle^{(2)}$. These vectors have the
properties analogous to \eqref{Tjj}
\be{vac-vec}
T^{(k)}_{ij}(u)|0\rangle^{(k)}=0,\quad i>j,\quad T^{(k)}_{ii}(u)|0\rangle^{(k)}=\lambda^{(k)}_i(u)|0\rangle^{(k)},\quad k=1,2.
\ee
Similarly to \eqref{ratios} we introduce ratios
 \be{ratios-part}
 r_1^{(k)}(w)=\frac{\lambda_1^{(k)}(w)}{\lambda_2^{(k)}(w)}, \qquad  r_3^{(k)}(w)=\frac{\lambda_3^{(k)}(w)}{\lambda_2^{(k)}(w)}, \qquad k=1,2.
 \ee
Due to the normalization $\lambda_2(u)=1$ we can always set $\lambda^{(k)}_2(u)=1$. Below we also extend  convention \eqref{SH-prod} to the products
of functions \eqref{ratios-part}.

One can construct for every partial monodromy matrix $T^{(k)}(u)$ the corresponding partial Bethe vectors $\mathbb{B}_{a,b}^{(k)}(\bla;\bmu)$.
They are given by equation \eqref{BV-expl}, where one should replace all $T_{ij}(u)$ by $T^{(k)}_{ij}(u)$ and $|0\rangle$ by $|0\rangle^{(k)}$.
The main problem considered in the framework of the composite model is to express total Bethe vectors $\mathbb{B}_{a,b}(\bla;\bmu)$ in terms of partial
$\mathbb{B}_{a,b}^{(k)}(\bla;\bmu)$. This problem was solved in \cite{IzeK84} for $GL(2)$-based models. More general case of $GL(N)$-invariant
models was considered in \cite{TarVar93,BeKhP07}. Particular case of $GL(3)$-invariant
models was studied in \cite{PakRS15b}, where the following representation was found:
\be{BV-BV}
\mathbb{B}_{a,b}(\bla;\bmu)=\sum r_{1}^{(2)}(\bla^{(1)}) r_{3}^{(1)}(\bmu^{(2)})\frac{f(\bla^{(2)},\bla^{(1)})f(\bmu^{(2)},\bmu^{(1)})}{f(\bmu^{(2)},\bla^{(1)})}\;
\mathbb{B}_{a_1,b_1}^{(1)}(\bla^{(1)};\bmu^{(1)}) \mathbb{B}_{a_2,b_2}^{(2)}(\bla^{(2)};\bmu^{(2)}).
\ee
Here the sum is taken over all possible partitions $\bla\Rightarrow\{\bla^{(1)},\bla^{(2)}\}$
and $\bmu\Rightarrow\{\bmu^{(1)},\bmu^{(2)}\}$. The cardinalities of the subsets are shown by the subscripts of the
partial Bethe vectors.

Similarly we can define a multi-composite model, where the original interval is divided into $M>2$ intervals
\be{T-TTT}
T(u)=T^{(M)}(u)\dots T^{(1)}(u).
\ee
For each
of these intervals we can define partial Bethe vectors $\mathbb{B}_{a_j,b_j}^{(j)}$. Then the total Bethe vector can be expressed
in terms of the partial ones as follows
\be{BV-BV-mult}
\mathbb{B}_{a,b}(\bla;\bmu)=\sum \prod_{1\le k<j\le M}\left\{ r_{1}^{(j)}(\bla^{(k)})r_{3}^{(k)}(\bmu^{(j)})
\frac{f(\bla^{(j)},\bla^{(k)})f(\bmu^{(j)},\bmu^{(k)})}
{f(\bmu^{(j)},\bla^{(k)})}\right\}\;
\prod_{j=1}^M\mathbb{B}_{a_j,b_j}^{(j)}(\bla^{(j)};\bmu^{(j)}).
\ee
Here the functions $r_{1}^{(j)}(u)$ and $r_{3}^{(j)}(v)$ are vacuum eigenvalues of the operators $T_{11}^{(j)}(u)$ and $T_{33}^{(j)}(v)$ respectively.
The sum in \eqref{BV-BV-mult} is taken over all possible partitions
\be{subsets}
\begin{aligned}
&\bla\Rightarrow\{\bla^{(1)},\dots,\bla^{(M)}\}, \qquad &\#\bla^{(j)}=a_j,\qquad& a_1+\dots+a_M=a,\\
&\bmu\Rightarrow\{\bmu^{(1)},\dots,\bmu^{(M)}\},\qquad & \#\bmu^{(j)}=b_j,\qquad& b_1+\dots+b_M=b.
\end{aligned}
\ee

It is important that the number $M$ of the partial monodromy matrices  is not related to the cardinalities of the Bethe parameters $a$ and $b$.
In particular, we can have $M>a$ and $M>b$. In this case some of numbers $a_j$ and $b_j$ are equal to zero, that is,
the corresponding subsets are empty.

Equation \eqref{BV-BV-mult} can be easily proved by induction over $M$. Indeed, assuming that it is valid for $M-1$ partial monodromy matrices
we  apply \eqref{BV-BV} to the partial Bethe vector $\mathbb{B}_{a_{M-1},b_{M-1}}^{(M-1)}(\bla^{(M-1)};\bmu^{(M-1)})$. This
immediately gives \eqref{BV-BV-mult} for $M$ partial monodromy matrices.

In particular cases $a=0$ or $b=0$ we reproduce known formulas for Bethe vectors in $GL(2)$-invariant multi-composite model \cite{IzeKR86,KorBIL93}
For instance,
\be{BV-mult-GL2}
\mathbb{B}_{a,0}(\bla,\emptyset)\equiv \mathbb{B}_{a}(\bla)=\sum \prod_{1\le k<j\le M}\left\{ r_{1}^{(j)}(\bla^{(k)})
f(\bla^{(j)},\bla^{(k)}\right\}\;
\prod_{j=1}^M\mathbb{B}_{a_j}^{(j)}(\bla^{(j)}).
\ee
The multi-composite model is a convenient way to express the Bethe vectors in terms of local operators.
In the next section, we discuss the method in more detail.

\subsection{Bethe vectors in the $SU(2)$ $XXX$ chain\label{S-BV-XXX}}

As a first application of the multi-composite model we construct Bethe vectors of the $SU(2)$ inhomogeneous $XXX$ chain.
This result will be used in section~\ref{S-TC-BG} for description of Bethe vectors of TCBG model.

Consider an inhomogeneous $XXX$ chain consisting of $M$ sites. This model has a $2\times 2$ monodromy matrix $T^{(xxx)}(u)$,
therefore Bethe vectors are parameterized by only one set of the Bethe parameters, say $\bla$. Respectively, considering the
multi-composite model one should use
\eqref{BV-mult-GL2}.

The monodromy matrix is defined as a product of local $L$-operators
\be{T-Lxxx}
T^{(xxx)}(u)=L_M^{(xxx)}(u-\xi_M)\dots L_1^{(xxx)}(u-\xi_1),
\ee
where $\xi_k$ are inhomogeneities and
\begin{equation}\label{L-opXXX}
L_n^{(xxx)}(u)=\frac1{u}\begin{pmatrix}
u+\frac c2(1+\sigma^z_n)& c\;\sigma^-_n \\
c\;\sigma^+_n &
u+\frac c2(1-\sigma^z_n)
\end{pmatrix}.
\end{equation}
Here $\sigma^z_n$ and $\sigma^\pm_n$ are spin-$1/2$ operators acting in the $n$-th site of the chain. They are given by the standard
Pauli matrices acting in the $n$-th copy of the tensor product $\bigl(\mathbb{C}^{2}\bigr)^{\otimes M}$. The pseudovacuum vector is the state
with all spins up
\be{vac-XXX}
|\tilde 0\rangle = \left(\begin{smallmatrix} 1\\0\end{smallmatrix}\right)_M\otimes\dots\otimes\left(\begin{smallmatrix} 1\\0\end{smallmatrix}\right)_1.
\ee
Bethe vectors with $a$ spins down and $M-a$ spins up have the form
\be{BV-xxx-Om}
\mathbb{B}^{(xxx)}_{a}(\bla)=
\sum_{M\ge j_a>\dots>j_1\ge 1}\Omega^{(a,M)}_{j_1,\dots,j_a}(\bla;\bar\xi)\prod_{m=1}^a\sigma^-_{j_m}|\tilde 0\rangle.
\ee
where $\Omega^{(a,M)}_{j_1,\dots,j_a}(\bla;\bar\xi)$ are coefficients depending on the Bethe parameters $\bla$ and inhomogeneities $\bar\xi$.
Let us find these coefficients explicitly.

Consider a multi-composite model with $M$ partial monodromy matrices $T^{(j)}$. It means that every  $T^{(j)}$ coincides with the
$L$-operator $L_j(u-\xi_j)$. Then every  partial Bethe vectors $\mathbb{B}_{a_j}^{(j)}(\bla^{(j)})$ in \eqref{BV-mult-GL2} corresponds to the $j$-th site of the chain, therefore due to \eqref{L-opXXX} we obtain
\be{PBV-xxx}
\mathbb{B}_{a_j}^{(j)}(\bla^{(j)})=g(\bla^{(j)},\xi_j)     \left(\sigma^-_j\right)^{a_j}\left(\begin{smallmatrix} 1\\0\end{smallmatrix}\right)_j.
\ee
Obviously $\mathbb{B}_{a_j}^{(j)}$ vanishes if $a_j>1$, because $\bigl(\sigma^-_j\bigr)^2=0$. Thus, we conclude that $a_j\le 1$ and the subsets $\bla^{(j)}$ are either empty or they consist of one element. Let subsets $\bla^{(j_k)}$ ($k=1,\dots,a$) corresponding to the lattice sites $j_1,\dots, j_a$ contain one element $u_k$, while other subsets are empty. Then the sum over partitions of the set $\bla$ turns into the sum over permutations in $\bla$ and the sum over the lattice sites $j_1,\dots, j_a$ with the restriction $j_a>\dots>j_1$.

It is easy to see that
\be{eig-XXX-vac}
\frac{u-\xi_j+\frac c2(1+\sigma^z_j)}{u-\xi_j}\left(\begin{smallmatrix} 1\\0\end{smallmatrix}\right)_j=
f(u,\xi_j)\left(\begin{smallmatrix} 1\\0\end{smallmatrix}\right)_j,
\qquad
\frac{u-\xi_j+\frac c2(1-\sigma^z_j)}{u-\xi_j}\left(\begin{smallmatrix} 1\\0\end{smallmatrix}\right)_j=
\left(\begin{smallmatrix} 1\\0\end{smallmatrix}\right)_j,
\ee
and  thus,
\be{r1}
r_{1}^{(j)}(u)=f(u,\xi_j).
\ee
Then equation \eqref{BV-mult-GL2} takes the form
\begin{equation}\label{BV-xxx-mult-2}
\mathbb{B}^{(xxx)}_{a}(\bla)=\Sym_{\bar u} \prod_{1\le k<j\le a}f(u_j,u_k)
\sum_{M\ge j_a>\dots>j_1\ge 1}
\prod_{k=1}^a  \left[\Bigl(\prod\limits_{m=j_k+1}^M f(u_k,\xi_m)\Bigr) g(u_k,\xi_{j_k})\sigma^-_{j_k}\right]|\tilde 0\rangle,
\end{equation}
where the  symbol $\Sym$ means symmetrization (i.e. the sum over permutations) over the set indicated by the subscript.
The symmetrization in \eqref{BV-xxx-mult-2} acts on all the expression depending on $\bla$.
Comparing \eqref{BV-xxx-mult-2} with \eqref{BV-xxx-Om} we see that

\begin{equation}\label{Omega-expl}
\Omega^{(a,M)}_{j_1,\dots,j_a}(\bla;\bar\xi)=\Sym_{\bar u} \prod_{1\le k<j\le a}f(u_j,u_k)
\prod_{k=1}^a  \left[\Bigl(\prod\limits_{m=j_k+1}^M f(u_k,\xi_m)\Bigr) g(u_k,\xi_{j_k})\right].
\end{equation}
In the homogeneous limit $\xi_k=c/2$ this expression coincides with the amplitude of the Bethe vector
in the coordinate Bethe ansatz representation (see \cite{Gaud83}).

\section{Two-component Bose gas\label{S-TC-BG}}

We consider the TCBG model on a finite interval $[0,L]$ with periodic boundary conditions. In the second quantized
form the Hamiltonian  has the form
 \be{HamQ}
 H=\int_0^L\left(\partial_x\Psi^\dagger_\alpha\partial_x\Psi_\alpha+\cc\Psi^\dagger_\alpha\Psi^\dagger_\beta\Psi_\beta\Psi_\alpha\right)\,dx,
 \ee
where $\cc>0$ is a coupling constant, $\alpha,\beta=1,2$ and the summation over repeated subscripts is assumed.
Bose fields $\Psi_\alpha(x)$ and $\Psi_\alpha^\dagger(x)$ satisfy canonical commutation relations
\be{com-con}
[\Psi_\alpha(x),\Psi_\beta^\dagger(y)]=\delta_{\alpha\beta}\delta(x-y).
\ee
The coupling constant $\cc$ is related to the constant $c$ in \eqref{R-mat} by $\cc=ic$.

The basis in the Fock space of the model is constructed by acting with operators
$\Psi_\alpha^\dagger(x)$ onto the Fock vacuum $|0\rangle$ defined as
\be{vac-def}
\Psi_\alpha(x)|0\rangle=0,\qquad \langle 0|\Psi_\alpha^\dagger(x)=0, \qquad \langle 0|0\rangle=1.
\ee
Observe that in the case of the TCBG model the pseudovacuum vector \eqref{Tjj} coincides with the
Fock vacuum $|0\rangle$, therefore we use the same notation for them.

The spectral problem for the TCBG model was solved in \cite{Yang67} (see also \cite{Sath68,Gaud83}).
The Hamiltonian eigenvectors can be found in two steps. Using the terminology of the algebraic Bethe ansatz one can say that at the first stage one should construct
a generic Bethe vector $\mathbb{B}_{a,b}(\bla;\bmu)$.  In the TCBG model Bethe vectors exist for  $a\le b$. They have the following form\footnote{%
Here and below we do not take care about eigenvectors  normalization in all formulas for them.}:
\begin{multline}\label{BV-coord-1c}
\mathbb{B}_{a,b}(\bla;\bmu)=
 \sum_{b\ge k_a>\dots>k_1\ge 1}\;\;
 \int\limits_{\mathcal{D}}
 \,dz_1\dots dz_b\;\chi_{k_1,\dots,k_a}(z_1,\dots,z_b|\bla,\bmu)\\
 \times
\prod_{m=1}^a\Psi^\dagger_1(z_{k_m}) \prod_{\substack{l=1\\ \ell\notin \{k_1,\dots,k_m\}}}^b\Psi^\dagger_2(z_\ell)|0\rangle.
\end{multline}
Here the integration domain is $\mathcal{D}=L>z_b>\dots>z_1>0$. In this domain the wave function $\chi_{k_1,\dots,k_a}(z_1,\dots,z_b|\bla,\bmu)$ has the
form
\begin{equation}\label{BV-coord-2c}
\chi_{k_1,\dots,k_a}(z_1,\dots,z_b|\bla,\bmu)=\Sym_{\bmu}\Omega^{(a,b)}_{k_1,\dots,k_a}(\bla;\bmu+c)\prod_{b\ge j>k\ge 1} f(v_j,v_k)
\prod_{k=1}^b  e^{iz_kv_k}\Bigr|_{c=-i\cc}\;,
\end{equation}
where  the coefficients $\Omega^{(a,b)}_{k_1,\dots,k_a}(\bla;\bmu+c)$ are given by \eqref{Omega-expl}.

Generic Bethe vector \eqref{BV-coord-1c} becomes  an eigenvector of the Hamiltonian \eqref{HamQ} if the parameters $\bla$ and $\bmu$ satisfy
the system of Bethe equations \eqref{AEigenS-1}. In the TCBG model it has the following form \cite{Yang67}:
\be{BE-0}
\begin{aligned}
e^{iLv_{j}}&=\prod_{\substack{k=1\\k\ne j}}^b\frac{v_{j}-v_{k}+i\cc}{v_{j}-v_{k}-i\cc}
\prod_{\ell=1}^a\frac{u_\ell-v_{j}+i\cc}{u_\ell-v_{j}}, \qquad j=1,\dots,b,\\
1&=\prod_{\substack{\ell=1\\ \ell\ne j}}^a\frac{u_{i}-u_\ell-i\cc}{u_{i}-u_\ell+i\cc}
\prod_{k=1}^b\frac{v_k-u_{i}-i\cc}{v_k-u_{i}},\qquad i=1,\dots,a.
\end{aligned}
\ee
Comparing this system with \eqref{AEigenS-1} we conclude that in the TCBG model $r_1(u)=1$ and $r_3(u)=e^{iLu}$.

\section{Lattice two-component Bose gas\label{S-LTC-BG}}

Quantum systems describing by the $GL(3)$-invariant $R$-matrix \eqref{R-mat} were considered in \cite{KulRes82}.
There a prototype of a lattice $L$-operator of the TCBG model was found. It has the following form:
\be{L-kulresh}
L^{(a)}(u)=u\mathbf{1}+p,
\ee
where
\be{p-kulresh}
p=\begin{pmatrix}a^\dagger_1a_1& a^\dagger_1a_2& ia^\dagger_1\sqrt{m+\rho}\\
a^\dagger_2a_1& a^\dagger_2a_2& ia^\dagger_2\sqrt{m+\rho}\\
i\sqrt{m+\rho}\;a_1& i\sqrt{m+\rho}\;a_2& -m-\rho
\end{pmatrix}.
\ee
Here $m$ is an arbitrary complex number and $\rho=a^\dagger_1a_1+a^\dagger_2a_2$. The operators $a_k$ and $a^\dagger_k$ ($k=1,2$)
act in a Fock space with the Fock vacuum $|0\rangle$: $a_k|0\rangle=0$. They have standard commutation
relations of the Heisenberg algebra $[a_i,a^\dagger_k]=\delta_{ik}$.

The $L$-operator \eqref{L-kulresh} satisfies the algebra \eqref{RTT} with
$R$-matrix \eqref{R-mat} at $c=-1$. Basing on the $L$-operator \eqref{L-kulresh} one can construct a quantum system of discrete bosons.
In order to obtain a continuous quantum system one should make several transforms of \eqref{L-kulresh}.
First, we introduce operators
\be{psi-a}
\psi_k=\Delta^{-1/2}a_k, \qquad  \psi^\dagger_k=\Delta^{-1/2}a^\dagger_k, \qquad k=1,2,
\ee
so that
\be{com-rel-psi}
[\psi_j, \psi^\dagger_k]=\frac{\delta_{jk}}\Delta.
\ee
In these formulas $\Delta$ is a lattice interval. Setting $m=\frac{4}{\cc\Delta}$ we introduce a new $L$-operator as
\be{IK-RK}
L(u)= \frac{\cc\Delta}2\;L^{(a)}\left(\frac{u+2i/\Delta}{i\cc}\right)\cdot J\;,
\ee
where $J=\diag(1,1,-1)$. Obviously $L(u)$ satisfies the $RTT$-relation \eqref{RTT} with
$R$-matrix \eqref{R-mat} at $c=-i\cc$.

The last transformation is to make $N$ copies $L_n$  ($n=1,\dots,N$) of $L$-operator \eqref{IK-RK} by changing
$\psi_k\to \psi_k(n)$ and $\psi^\dagger_k\to \psi^\dagger_k(n)$ with
\be{com-rel-psin}
[\psi_j(n), \psi^\dagger_k(m)]=\frac{\delta_{jk}\delta_{nm}}\Delta.
\ee
The operators $\psi_k(n)$ and $\psi^\dagger_k(n)$ are lattice approximations of the Bose fields $\Psi_k(x)$ and  $\Psi^\dagger_k(x)$. Indeed,
let us divide the interval $[0,L]$ into $N$ sites of the length $\Delta$. Setting $x_n=n\Delta$ and
\be{lat-con}
\psi_k(n)=\frac1\Delta\int_{x_{n-1}}^{x_n}\Psi_k(x)\,dx, \qquad \psi^\dagger_k(n)=\frac1\Delta\int_{x_{n-1}}^{x_n}\Psi^\dagger_k(x)\,dx,
\ee
we reproduce commutation relations \eqref{com-rel-psin}. On the other hand, in the limit $\Delta\to 0$ the operators \eqref{lat-con}
obviously turn into the Bose fields\footnote{Here and below limits of operator-valued expressions should be understood in the weak sense.}
 $\Psi_k(x)$ and $\Psi^\dagger_k(x)$.

Now we can define a monodromy matrix in a standard way
\be{T-L}
T(u)=L_N(u)\dots L_1(u),
\ee
where
\begin{equation}\label{L-op}
L_n(u)=\frac1{\mathcal{N}}\begin{pmatrix}
1-\frac{iu\Delta}2+\frac{\cc\Delta^2}2\psi_{1}^\dagger(n)\psi_{1}(n)& \frac{\cc\Delta^2}2\psi_{1}^\dagger(n)\psi_{2}(n)&
-i\Delta\psi_{1}^\dagger(n) Q_n\\
\frac{\cc\Delta^2}2\psi_{2}^\dagger(n)\psi_{1}(n)&1-\frac{iu\Delta}2+ \frac{\cc\Delta^2}2\psi_{2}^\dagger(n)\psi_{2}(n)&
-i\Delta\psi_{2}^\dagger(n) Q_n\\
i\Delta Q_n\psi_{1}(n)&
i\Delta Q_n\psi_{2}(n)&
1+\frac{iu\Delta}2+\frac{\cc\Delta^2}2\hat\rho_n
\end{pmatrix},
\end{equation}
and
\be{rho}
\mathcal{N}=\left(1-\frac{iu\Delta}2\right), \qquad \hat\rho_n=\psi_{1}^\dagger(n)\psi_{1}(n)+\psi_{2}^\dagger(n)\psi_{2}(n),\qquad Q_n=\left(\cc+\frac{\cc^2\Delta^2}4\hat\rho_n\right)^{1/2}.
\ee
The normalization factor $\mathcal{N}$ in \eqref{L-op} is used in order to satisfy the condition $\lambda_2(u)=1$.

{\sl Remark.} We write the number of the lattice site $n$ as the argument of the operators $\psi_i$ and $\psi_i^\dagger$.
Traditionally this number is written as subscript of $\psi_i$ and $\psi_i^\dagger$, but in the case of the TCBG model it is not
convenient.

$L$-operator \eqref{L-op} is a natural generalization of a $2\times 2$ $L$-operator found in \cite{IzeK81} for the lattice model
of one-component  bosons:
\begin{equation}\label{L-op22}
\tilde L_n(u)=\frac1{\mathcal{N}}\begin{pmatrix}
1-\frac{iu\Delta}2+ \frac{\cc\Delta^2}2\psi^\dagger(n)\psi(n)&
-i\Delta\psi^\dagger(n) Q_n\\
i\Delta Q_n\psi(n)&
1+\frac{iu\Delta}2+\frac{\cc\Delta^2}2\psi^\dagger(n)\psi(n)
\end{pmatrix}.
\end{equation}
It is easy to see that $L$-operator \eqref{L-op22} is the right-lower $2\times 2$ minor of the matrix \eqref{L-op} with the
identification $\psi_1(n)\equiv 0$, $\psi_2(n)\equiv \psi(n)$. It was shown by different methods in \cite{IzeK82,TarTF83,BogK86} that
in the continuous limit $\Delta\to 0$ the $L$-operator \eqref{L-op22} describes the model of one-dimensional bosons with $\delta$-function interaction.
We have to solve an analogous problem: to check that in the continuous limit  the model with the monodromy matrix \eqref{T-L}
and $L$-operator \eqref{L-op} does describe the TCBG model. For this purpose  we will find Bethe vectors of the lattice
model \eqref{T-L} and will show that they coincide with the states \eqref{BV-coord-2c} in the continuous limit.

Let us point out sever properties of the $L$-operator \eqref{L-op}. It is easy to see that
\be{prop-L}
\begin{aligned}
&\bigl(L_n(u)\bigr)_{11}|0\rangle=\bigl(L_n(u)\bigr)_{22}|0\rangle=|0\rangle,\qquad \bigl(L_n(u)\bigr)_{33}|0\rangle=
\rr(u)\;|0\rangle;\\
&\bigl(L_n(u)\bigr)_{12}|0\rangle=0,\qquad \langle0|\bigl(L_n(u)\bigr)_{21}=0,
\end{aligned}
\ee
where
\be{r0}
\rr(u)=\left(\frac{1+\frac{iu\Delta}2}{1-\frac{iu\Delta}2}\right).
\ee
From these properties we easily find
\be{prop-T}
\begin{aligned}
&r_1(u)=1,\qquad &r_3(u)=\rr^N(u);\\
&T_{12}(u)|0\rangle=0,\qquad &\langle0|T_{21}(u)=0.
\end{aligned}
\ee
Note that in fact the condition $r_1(u)=1$ implies the actions of $T_{12}(u)$ and $T_{21}(u)$ in the second line of \eqref{prop-T}. Indeed,
we have from the $RTT$-relation \eqref{RTT}
\be{RTT-1221}
[T_{21}(v),T_{12}(u)]=g(v,u)\bigl(T_{11}(u)T_{22}(v)-T_{11}(v)T_{22}(u)\bigr).
\ee
Applying this equation, for example, to the vector $|0\rangle$ and using $r_1(u)=1$ we obtain
\begin{multline}\label{RTT-1221-0}
[T_{21}(v),T_{12}(u)]|0\rangle=T_{21}(v)T_{12}(u)|0\rangle\\
=g(v,u)\bigl(T_{11}(u)T_{22}(v)-T_{11}(v)T_{22}(u)\bigr)|0\rangle
=\bigl(r_{1}(u)-r_{1}(v)\bigr)|0\rangle=0.
\end{multline}
Similarly, acting with \eqref{RTT-1221} on $\langle0|$ we obtain $\langle0|T_{21}(u)=0$.

The property $T_{12}(u)|0\rangle=0$ leads to a simplification of the explicit formula for the Bethe vector \eqref{BV-expl}.
Obviously, in this case we should consider only such partitions of the set $\bla$ that $\bla^{(2)}=\emptyset$, and $\bla^{(1)}=\bla$.
Then \eqref{BV-expl} turns into
\be{BV-expl-BG}
\mathbb{B}_{a,b}(\bla;\bmu) =\sum \frac{\Izer_{a}(\bmu^{(1)}|\bla)}{f(\bmu,\bla)}
f(\bmu^{(2)},\bmu^{(1)})\,
T_{13}(\bmu^{(1)})T_{23}(\bmu^{(2)})|0\rangle.
\ee
Here the sum is taken over partitions of the only one set  $\bar v\Rightarrow\{\bar v^{(1)},\bar v^{(2)}\}$ with a restriction $\#\bar v^{(1)}=a$.
The last restriction evidently can be satisfied if and only if $a\le b$. Hence, if $a>b$, then $\mathbb{B}_{a,b}(\bla;\bmu)=0$. In particular,
\be{BV-expl-BG11}
\mathbb{B}_{0,1}(\emptyset;v) = T_{23}(v)|0\rangle,  \qquad \mathbb{B}_{1,1}(u;v) = \frac{g(v,u)}{f(v,u)}\;T_{13}(v)|0\rangle.
\ee

To conclude this section  we give two formulas concerning the continuous limit $\Delta\to0$. The first formula
gives the limit of powers of the function $\rr(u)$
\be{r0-n}
\lim_{\Delta\to 0}\rr^n(u)=\lim_{\Delta\to 0}\left(\frac{1+\frac{iu\Delta}2}{1-\frac{iu\Delta}2}\right)^{x_n/\Delta}=e^{iux_n}.
\ee
The second formula describes  a typical procedure of taking continuous limit of sums over the lattice sites.
Let $\Phi(x)$ be an integrable function on the interval $[0,L]$. Then
\be{sum-int}
\Delta\sum_{j=1}^N\Phi(x_j)\psi^\dagger(j)=
\sum_{j=1}^N\Phi(x_j)\int_{x_{j-1}}^{x_j}\Psi^\dagger_k(x)\,dx \longrightarrow \int_{0}^{L}\Phi(x)\Psi^\dagger_k(x)\,dx,\qquad \Delta\to0,
\ee
and we recall that all limits of operator-valued expressions are understood in a weak sense.
Thus we can formulate a general rule: a sum over the lattice sites multiplied by $\Delta$ turns into an integral in the continuous limit.
It is easy to see that if we have an $m$-fold sum over the lattice sites multiplied by $\Delta^m$, then it turns into an $m$-fold integral in the continuous limit.

\section{Bethe vectors in terms of local operators\label{S-BV-LO}}

Consider a multi-composite model with the total monodromy matrix \eqref{T-L}. Let the number $M$ of the partial monodromy matrices  coincides with the
number $N$ of the lattice sites.  Then every partial monodromy matrix $T^{(n)}(u)$ is the $L$-operator $L_n(u)$ \eqref{L-op}. Respectively we have
\be{r3}
r_{1}^{(k)}(u)= 1, \qquad r_{3}^{(k)}(v)= \rr(v).
\ee
The formula for the total Bethe vector \eqref{BV-BV-mult} takes the form
\be{BV-BV-mult-p}
\mathbb{B}_{a,b}(\bla;\bmu)=\sum \prod_{j=1}^N \rr^{j-1}(\bmu^{(j)})
\prod_{1\le k<j\le N}\frac{f(\bla^{(j)},\bla^{(k)})f(\bmu^{(j)},\bmu^{(k)})}
{f(\bmu^{(j)},\bla^{(k)})}\;
\prod_{j=1}^N\mathbb{B}_{a_j,b_j}^{(j)}(\bla^{(j)};\bmu^{(j)}).
\ee
This is the main formula that we shall use. But before applying this formula to the TCBG model it is useful to look
how it works for a more simple example of the one-component Bose gas.

\subsection{One-component Bose gas\label{SS-OC-BG}}

The $L$-operator of the one-component Bose gas is given by \eqref{L-op22}, however for the construction of Bethe vectors we need to know this
$L$ -operator only up to terms of order $\Delta $:
\begin{equation}\label{LIK-op}
\tilde L_n(u)=\begin{pmatrix}
1-\frac{iu\Delta}2&
-i\Delta\sqrt{\cc}\psi^\dagger(n) \\
i\Delta \sqrt{\cc} \psi(n)&
1+\frac{iu\Delta}2
\end{pmatrix} + O(\Delta^2).
\end{equation}
Recall that there we have set $\psi_2(n)\equiv \psi(n)$, $\psi_1(n)\equiv 0$
and similarly for $\psi^\dagger_k(n)$.
In the continuous limit these operators respectively turns into Bose fields $\Psi(x)$ and $\Psi^\dagger(x)$.

Bethe vectors of the one-component Bose gas correspond to the particular case of $\mathbb{B}_{a,b}(\bla;\bmu)$ at $a=0$ and
$\bla=\emptyset$. Then the formula \eqref{BV-BV-mult-p} takes the form
\be{BV-BV-mult1}
\mathbb{B}_{0,b}(\emptyset;\bmu)\equiv \mathbb{B}_{b}(\bmu)=\sum \prod_{j=1}^N \rr^{j-1}(\bmu^{(j)})
\prod_{1\le k<j\le N}f(\bmu^{(j)},\bmu^{(k)})\;
\prod_{j=1}^N\mathbb{B}_{b_j}^{(j)}(\bmu^{(j)}).
\ee
A partial Bethe vector in the site $j$ is
\be{BV-ex}
\mathbb{B}_{b_j}^{(j)}(\bmu^{(j)})= \bigl(-i\Delta\sqrt{\cc}\psi^\dagger(j)\bigr)^{b_j}|0\rangle,
\ee
where corrections of the order $O(\Delta^{b_j+1})$ are neglected.

{\sl Remark.} Recall that in the multi-composite model the total pseudovacuum vector $|0\rangle$ is equal to the tensor product
of the partial pseudovacuum vectors $|0\rangle^{(j)}$ ($j=1,\dots,N$). However,  in the case of the one-component Bose gas we can assume that
all operators $\psi(j)$ and $\psi^\dagger(j)$ act in the same Fock space. Obviously, due to commutativity of
$\psi(j)$ and $\psi^\dagger(k)$ at $j\ne k$ such the formulation is equivalent to the original one. In the case of the TCBG model we will use the same treatment
of the multi-composite model.

Consider an example $b=2$. Then we have two possibilities.

\begin{itemize}
\item There exists one $b_j$ such that $b_j=2$, while all other $b_\ell=0$. Then the subset $\bmu^{(j)}$ coincides with the
original set $\{v_1,v_2\}$, while all other subsets $\bmu^{(\ell)}$ are empty.

\item There exist two $b_j$ and $b_k$ such that $b_j=b_k=1$, while all other $b_\ell=0$. Then the subsets $\bmu^{(j)}$ and
$\bmu^{(k)}$ consist of one element (say, $\bmu^{(j)}=v_2$ and $\bmu^{(k)}=v_1$ or vice versa). All other subsets $\bmu^{(\ell)}$ are empty.

\end{itemize}

Consider the first case. We denote the corresponding contribution to the Bethe vector by $\mathbb{B}_{2,\emptyset}$. Then
\be{B-20}
\mathbb{B}_{2,\emptyset}=-\cc\Delta^2\sum_{j=1}^N \bigl(r_0(v_1)r_0(v_2)\bigr)^{j-1}\bigl(\psi^\dagger(j)\bigr)^{2}|0\rangle,
\ee
and due to \eqref{r0-n} we obtain
\be{B-20-1}
\mathbb{B}_{2,\emptyset}=-\cc\Delta^2\sum_{j=1}^N e^{ix_j(v_1+v_2)}\bigl(\psi^\dagger(j)\bigr)^{2}|0\rangle.
\ee
This sum goes to zero, because it has the coefficient $\Delta^2$. Indeed, due to \eqref{sum-int} we have
%
\be{fin-lim}
\Delta^2\sum_{j=1}^N e^{ix_j(v_1+v_2)}\bigl(\psi^\dagger(j)\bigr)^{2}|0\rangle \longrightarrow
\Delta\int_0^L e^{ix(v_1+v_2)}\bigl(\Psi^\dagger(x)\bigr)^{2}\,dx|0\rangle\longrightarrow 0,\qquad \Delta\to 0.
\ee

It remains to consider the second case. We denote the corresponding contribution to the Bethe vector by $\mathbb{B}_{1,1,\emptyset}$. Then
\be{B-110}
\mathbb{B}_{1,1,\emptyset}=-\cc\Delta^2\Sym_{\bmu}\sum_{1\le k<j\le N} r_0^{j-1}(v_2)r_0^{k-1}(v_1)f(v_2,v_1)\psi^\dagger(j)\psi^\dagger(k)|0\rangle,
\ee
or due to \eqref{r0-n}
\be{B-110-1}
\mathbb{B}_{1,1,\emptyset}=-\cc\Delta^2\Sym_{\bmu}\sum_{1\le k<j\le N} e^{ix_kv_1+ix_jv_2}\;f(v_2,v_1)\psi^\dagger(j)\psi^\dagger(k)|0\rangle.
\ee

This time we have again the coefficient $\Delta^2$, but the sum is double. Therefore the limit is finite
\be{B-110-2}
\lim_{\Delta\to 0}\mathbb{B}_{1,1,\emptyset}=\mathbb{B}_{2}(\bmu)=-\cc\Sym_{\bmu}f(v_2,v_1)\int_0^L\,dx_2 \int_0^{x_2}\,dx_1 e^{ix_1v_1+ix_2v_2}\;\Psi^\dagger(x_2)\Psi^\dagger(x_1)|0\rangle.
\ee

It is clear from \eqref{BV-BV-mult1} and \eqref{BV-ex} that for general $b$ the Bethe vector $\mathbb{B}_{b}(\bmu)$
is proportional to $\Delta^b$. In the continuous limit this coefficient should be compensated. The only possible way
to obtain such the compensation is to have a $b$-fold sum over the lattice sites. Then $\Delta^b$ times $b$-fold sum gives a $b$-fold integral.
Hence, we should consider only such  partitions of the set $\bmu=\{v_1,\dots,v_b\}$ that reduce to $b$-fold sums over the lattice sites.
Obviously, these are such partitions in which there are  exactly $b$ nonempty subsets. In this case every such subset consists of only one variable.
Thus, actually we deal with the case already considered in section~\ref{S-BV-XXX}.
Therefore the sum over partitions reduces to the sum over the lattice sites and the sum over permutations, i.e. to the symmetrization over $\bmu$.

Thus, we obtain for general $b$
\be{B-gen}
\mathbb{B}_{b}(\bmu)=(-i\sqrt{\cc}\Delta)^b\Sym_{\bmu}\prod_{b\ge j>k\ge 1} f(v_j,v_k)\sum_{j_b>\dots>j_1}^N \prod_{k=1}^b \left(r_0^{j_k-1}(v_k)\psi^\dagger(j_k)\right)|0\rangle,
\ee
or partly taking continuous limit
\be{B-gen-1}
\mathbb{B}_{b}(\bmu)=(-i\sqrt{\cc}\Delta)^b\Sym_{\bmu}\prod_{b\ge j>k\ge 1} f(v_j,v_k)\sum_{j_b>\dots>j_1}^N \prod_{k=1}^b \left(e^{ix_{j_k}v_k}\psi^\dagger(j_k)\right)|0\rangle.
\ee
This $b$-fold sum over the lattice sites goes to a $b$-fold integral, and we finally arrive at
\be{B-gen-2}
\lim_{\Delta\to 0}\mathbb{B}_{b}(\bmu)=(-i\sqrt{\cc})^b\Sym_{\bmu}\prod_{b\ge j>k\ge 1} f(v_j,v_k)\int\limits_{\mathcal{D}}
 \,dx_1\dots dx_b\prod_{k=1}^b \left(e^{ix_kv_k}\Psi^\dagger(x_k)\right)|0\rangle,
\ee
where $\mathcal{D}= L>x_b>\dots>x_1>0$. Representation \eqref{B-gen-2} coincides with well know result for the Bethe vectors in the coordinate Bethe ansatz \cite{LiebL63,Lieb63,KorBIL93}.
Thus, we have constructed Bethe vectors in terms of the local Bose field $\Psi^\dagger(x)$ starting from the lattice $L$-operator
\eqref{LIK-op}.

\subsection{Two-component Bose gas}

The infinitesimal lattice $L$-operator of the TCBG model has the form \cite{Kul81}
\begin{equation}\label{L-op-sm}
L_n(u)=\begin{pmatrix}
1-\frac{iu\Delta}2& 0& -i\Delta\sqrt{\cc}\psi_{1}^\dagger(n) \\
0&1-\frac{iu\Delta}2& -i\Delta\sqrt{\cc}\psi_{2}^\dagger(n) \\
i\Delta\sqrt{\cc}\psi_{1}(n)& i\Delta\sqrt{\cc} \psi_{2}(n)& 1+\frac{iu\Delta}2
\end{pmatrix}+O(\Delta^2).
\end{equation}
We again consider a multi-composite model with the number of the partial monodromy matrices $T^{(n)}(u)$ equal to the number of the lattice sites.
Then every $T^{(n)}(u)$ of such the model coincides with the $L$-operator \eqref{L-op-sm}. First of all
let us find how Bethe vector depends on $\Delta$. In the TCBG model Bethe vectors are given by
\eqref{BV-expl-BG}. It is easy to see that the total number of creation operators $T_{13}$ and $T_{23}$ in \eqref{BV-expl-BG} is $b$. In the case of partial Bethe vectors $\mathbb{B}_{a_j,b_j}^{(j)}(\bla^{(j)};\bmu^{(j)})$
we have
\be{T23T13}
T_{13}(w)=-i\Delta\sqrt{\cc}\psi_{1}^\dagger(j),\qquad T_{23}(w)=-i\Delta\sqrt{\cc}\psi_{2}^\dagger(j).
\ee
Therefore
\be{B-part-D}
\mathbb{B}_{a_j,b_j}^{(j)}(\bla^{(j)};\bmu^{(j)})\sim \Delta^{b_j},\quad\text{and thus,}\quad \mathbb{B}_{a,b}(\bla;\bmu)\sim \Delta^{b}.
\ee

The Bethe vectors of the multi-composite TCBG model  are given by \eqref{BV-BV-mult-p}.
Using the same arguments as in the case of the one-component Bose gas we conclude that we should consider only such partitions of the set $\bmu$,
where we have exactly $b$ nonempty subsets consisting of one element. Then the sum over such partitions of the set $\bmu$ turns into the sum over permutations
of $\bmu$ and a $b$-fold sum over the lattice sites.

Consider now what happens with the partitions of the set $\bla$. In every partial Bethe vector $b_j\ge a_j$.
As we have shown above, all $b_j$ are equal either to zero or to one.
If $b_j=0$, then $a_j=0$. However if $b_j=1$, then either $a_j=1$ or $a_j=0$. In the first case we obtain a partial Bethe vector
of the form $B_{1,1}^{(j)}$, in the second case a partial Bethe vector of the form $B_{0,1}^{(j)}$. But since all nonempty subsets consist of exactly one element,  the sum over partitions of the set $\bla$
also turns into the sum over permutations in $\bla$ and a sum over the lattice sites where $a_j=1$.

Thus, the sum in \eqref{BV-BV-mult-p} is organised  as follows. First, we should choose a set $J$ consisting of $b$ numbers $J=\{j_1,\dots,j_b\}$. These are the
numbers of the lattice sites, where $b_{j_k}=1$. In all other sites $b_j=0$. We assume that the subset $\bmu^{(j_k)}$ consists of one element $v_k$. Taking symmetrization over $\bmu$ and the sum over all possible $j_k$ with the restriction $j_b>\dots>j_1$ we thus reproduce the sum over partitions of the set $\bmu$. More precisely, we reproduce only such partitions that eventually contribute into the continuous limit.

Up to this point everything is exactly as in the case of one-component bosons. Now we should take into account the partitions of the set $\bla$. For this
we should choose among the set $J=\{j_1,\dots,j_b\}$ a subset  of numbers $K$ consisting of $a$ elements: $K=\{j_{k_1},\dots, j_{k_a}\}$, $K\subset J$.
These are the numbers of the lattice sites, where $a_{j_{k_m}}=1$. In all other sites $a_j=0$.  We assume that the subset $\bla^{(j_{k_m})}$ consists of one element $u_m$.
Taking symmetrization over $\bla$ and the sum over all possible $j_{k_m}$ with the restriction $j_{k_a}>\dots>j_{k_1}$ we  reproduce the sum over partitions of the set $\bla$.

Summarizing all above we  recast \eqref{BV-BV-mult-p} as follows:
\begin{multline}\label{BV-pre-cont}
\mathbb{B}_{a,b}(\bla;\bmu)=\Sym_{\bmu,\bla}\prod_{b\ge j>k\ge 1} f(v_j,v_k)\prod_{a\ge j>k\ge 1} f(u_j,u_k)
\sum_{j_b>\dots>j_1}^N \sum_{\substack{j_{k_a}>\dots>j_{k_1}\\ j_{k_m}\in J}}\\
\times\prod_{m=1}^a\prod_{\ell=k_m+1}^b f^{-1}(v_\ell,u_m)\prod_{k=1}^b r_0^{j_k-1}(v_k)
\prod_{m=1}^a\mathbb{B}_{1,1}^{(j_{k_m})}(u_m;v_{k_m}) \prod_{ j_\ell\in J\setminus K}\mathbb{B}_{0,1}^{(j_\ell)}(\emptyset;v_\ell).
\end{multline}
Due to \eqref{BV-expl-BG11} and \eqref{L-op-sm} we find
\be{B11-B01}
\mathbb{B}_{0,1}^{(j)}(\emptyset;v)=-i\Delta\sqrt{\cc}\psi^\dagger_2(j)|0\rangle, \qquad
\mathbb{B}_{1,1}^{(j)}(u;v)=-i\Delta\sqrt{\cc}\;\frac{g(v,u)}{f(v,u)}\;\psi^\dagger_1(j)|0\rangle,
\ee
and using \eqref{r0-n} we obtain
\begin{multline}\label{BV-pre-cont-1}
\mathbb{B}_{a,b}(\bla;\bmu)=(-i\Delta\sqrt{\cc})^b\Sym_{\bmu,\bla}\prod_{b\ge j>k\ge 1} f(v_j,v_k)\prod_{a\ge j>k\ge 1} f(u_j,u_k)
\sum_{j_b>\dots>j_1}^N \sum_{\substack{j_{k_a}>\dots>j_{k_1}\\ j_{k_m}\in J}}\\
\times \prod_{m=1}^a\prod_{\ell=k_m+1}^b f^{-1}(v_\ell,u_m)
\prod_{k=1}^b e^{ix_kv_k}
\prod_{m=1}^a\frac{g(v_{k_m},u_m)}{f(v_{k_m},u_m)}\;\psi^\dagger_1(j_{k_m}) \prod_{ j_\ell\in J\setminus K}\psi^\dagger_2(j_\ell)|0\rangle.
\end{multline}
Using obvious properties of the functions $g(x,y)$ and $f(x,y)$
\be{obv-evid}
f(x,y+c)=\frac1{f(y,x)},\qquad g(x,y+c)=-\frac{g(y,x)}{f(y,x)},
\ee
we see that
\be{Om}
\Sym_{\bla}\prod_{a\ge j>k\ge 1} f(u_j,u_k)
\prod_{m=1}^a\Biggl\{\frac{g(v_{k_m},u_m)}{f(v_{k_m},u_m)}\prod_{\ell=k_m+1}^b \frac{1}{f(v_\ell,u_m)}\Biggr\}=
(-1)^a\Omega^{(a,b)}_{k_1,\dots,k_a}(\bla;\bmu+c),
\ee
where the coefficients $\Omega^{(a,b)}_{k_1,\dots,k_a}$ are given by \eqref{Omega-expl}.

Thus \eqref{BV-pre-cont-1} takes the form
\begin{multline}\label{BV-pre-cont-2}
\mathbb{B}_{a,b}(\bla;\bmu)=(-1)^a(-i\Delta\sqrt{\cc})^b\Sym_{\bmu}\prod_{b\ge j>k\ge 1} f(v_j,v_k)
\sum_{j_b>\dots>j_1}^N \sum_{\substack{j_{k_a}>\dots>j_{k_1}\\ j_{k_m}\in J}}\prod_{k=1}^b e^{ix_kv_k}\\
\times  \Omega^{(a,b)}_{k_1,\dots,k_a}(\bla;\bmu+c)
\;\prod_{m=1}^a\psi^\dagger_1(j_{k_m}) \prod_{ j_\ell\in J\setminus K}\psi^\dagger_2(j_\ell)|0\rangle,
\end{multline}
and it becomes clear that
in the continuous limit we arrive at \eqref{BV-coord-1c} up to a normalization factor.

\section{Representation of the monodromy matrix in terms of Bose fields\label{S-RMM-BF}}

In this section we derive explicit representations of the monodromy matrix elements $T_{ij}(u)$ in terms of the Bose fields. These representations have the
form of a formal power series in the coupling constant $\cc$. It is worth mentioning, however, that in a weak sense, these series are cut on an arbitrary
Bethe vector.

Let us present the infinitesimal $L$-operator \eqref{L-op-sm} as a block-matrix of the size $2\times 2$:
\begin{equation}\label{L-op-22}
L_n(u)=\begin{pmatrix}a&b_n\\c_n&d\end{pmatrix}+O(\Delta^2).
\end{equation}
Here $d=1+\frac{iu\Delta}2$, and $a$ is a $2\times 2$ matrix $a=(1-\frac{iu\Delta}2)\cdot\mathbf{1}$, where $\mathbf{1}$ is the identity
matrix of the size $2\times 2$. A two-component vector-column $b_n$ and two-component vector-row $c_n$ are
\be{bncn}
b_n= -i\Delta\sqrt{\cc}  \begin{pmatrix}\psi_{1}^\dagger(n)\\ \psi_{2}^\dagger(n)
\end{pmatrix}, \qquad
c_n= i\Delta\sqrt{\cc}  \begin{pmatrix}\psi_{1}(n);& \psi_{2}(n)
\end{pmatrix}.
\ee
It is convenient to separate the diagonal and anti-diagonal parts of the $L$-operator \eqref{L-op-22} as follows\footnote{%
Here and below we omit the terms $O(\Delta^2)$ as they  do not contribute to the continuous limit.}:
\begin{equation}\label{L-op-LW}
L_n(u)=\Lambda(u) +W_n,
\end{equation}
where
\begin{equation}\label{L-op-L-W}
\Lambda(u)=\begin{pmatrix}a&0\\0&d\end{pmatrix},\qquad W_n=\begin{pmatrix}0&b_n\\c_n&0\end{pmatrix}.
\end{equation}

Now representation \eqref{L-op-LW} should be substituted in \eqref{T-L} and then developed into the series in $W_n$.
Since the anti-diagonal part $W_n$ is proportional to $\sqrt{\cc}$, the monodromy matrix $T(u)$  becomes a polynomial in $\sqrt{\cc}$, that turns into an infinite power series in the continuous limit:
\be{T-power}
T(u)=\sum_{n=0}^\infty \cc^{n/2} T_n(u),
\ee
where
\be{Tnu}
\cc^{n/2}T_n(u)=\left(1-\frac{iu\Delta}2\right)^{-N}\sum_{N\ge k_n>\dots>k_1\ge 1}\Lambda^{N-k_n}W_{k_n}\Lambda^{k_n-k_{n-1}-1}
\cdots\Lambda^{k_2-k_1-1}W_{k_1}\Lambda^{k_1-1}.
\ee
It is clear from \eqref{Tnu} that the diagonal blocks of the monodromy matrix are series in integer powers of $\cc$, while
anti-diagonal blocks are series in half-integer powers of $\cc$.

Let
\be{tW}
\widetilde{W}_{k_i}=\Lambda^{-k_i}W_{k_i}\Lambda^{k_i-1}=\begin{pmatrix}0&\tilde b_{k_i}\\ \tilde c_{k_i}&0\end{pmatrix},
\ee
where
\be{tb-tc}
\tilde b_{k_i} =\frac{b_{k_i}}{1+\frac{iu\Delta}2}\left(\frac{1+\frac{iu\Delta}2}{1-\frac{iu\Delta}2} \right)^{k_i},
\qquad
\tilde c_{k_i} =\frac{c_{k_i}}{1-\frac{iu\Delta}2}\left(\frac{1-\frac{iu\Delta}2}{1+\frac{iu\Delta}2} \right)^{k_i}.
\ee
Then equation \eqref{Tnu} takes the form
\be{Tnu-1}
\cc^{n/2}T_n(u)=\left(1-\frac{iu\Delta}2\right)^{-N}\Lambda^{N}\sum_{N\ge k_n>\dots>k_1\ge 1}\widetilde{W}_{k_n}\widetilde{W}_{k_{n-1}}\cdots
\widetilde{W}_{k_1}.
\ee
Partly taking the continuous limit via \eqref{r0-n} we obtain
\be{LN}
\lim_{\Delta\to0}\left(1-\frac{iu\Delta}2\right)^{-N}\Lambda^{N}=\begin{pmatrix} \mathbf{1} &0\\0& e^{iuL} \end{pmatrix},
\ee
and
\be{tb-tc-cont}
\tilde b_{k_i} =b_{k_i}e^{iux_{k_i}},
\qquad
\tilde c_{k_i} =c_{k_i}e^{-iux_{k_i}}.
\ee

It is convenient to study the operators $T_n(u)$ separately for $n$ even and  $n$ odd. Let $n=2\ell$. The product of two matrices $\widetilde{W}_{k_i}$ and $\widetilde{W}_{k_{i-1}}$ gives a block-diagonal
matrix
\be{prod-two}
\widetilde{W}_{k_i}\widetilde{W}_{k_i-1}=\begin{pmatrix} \tilde b_{k_i}\tilde c_{k_{i-1}} &0\\0&\tilde c_{k_i}\tilde b_{k_{i-1}}\end{pmatrix}.
\ee
Hence, we obtain
\be{Tell-diag}
\cc^\ell T_{2\ell}(u)=\begin{pmatrix} A_\ell(u) &0\\0&D_\ell(u) \end{pmatrix},
\ee
where
\be{A-ell}
 A_\ell(u)=\sum_{N\ge k_{2\ell}>\dots>k_1\ge 1} \tilde b_{k_{2\ell}}\tilde c_{k_{2\ell-1}}
 \tilde b_{k_{2\ell-2}}\tilde c_{k_{2\ell-3}}\cdots \tilde b_{k_{2}}\tilde c_{k_{1}},
 \ee
and
\be{D-ell}
 D_\ell(u)=e^{iuL}\sum_{N\ge k_{2\ell}>\dots>k_1\ge 1} \tilde c_{k_{2\ell}}\tilde b_{k_{2\ell-1}}
 \tilde c_{k_{2\ell-2}}\tilde b_{k_{2\ell-3}}\cdots \tilde c_{k_{2}}\tilde b_{k_{1}}.
 \ee
Observe that all operators in these products commute, because they are from different lattice sites. Therefore
\be{Norm-form}
\begin{aligned}
\tilde b_{k_{2\ell}}\tilde c_{k_{2\ell-1}}
 \tilde b_{k_{2\ell-2}}\tilde c_{k_{2\ell-3}}\cdots \tilde b_{k_{2}}\tilde c_{k_{1}}&=\quad
 :\tilde b_{k_{2\ell}}\tilde c_{k_{2\ell-1}}
 \tilde b_{k_{2\ell-2}}\tilde c_{k_{2\ell-3}}\cdots \tilde b_{k_{2}}\tilde c_{k_{1}}:,\\
 \tilde c_{k_{2\ell}}\tilde b_{k_{2\ell-1}}
 \tilde c_{k_{2\ell-2}}\tilde b_{k_{2\ell-3}}\cdots \tilde c_{k_{2}}\tilde b_{k_{1}}&=\quad
 :\tilde c_{k_{2\ell}}\tilde b_{k_{2\ell-1}}
 \tilde c_{k_{2\ell-2}}\tilde b_{k_{2\ell-3}}\cdots \tilde c_{k_{2}}\tilde b_{k_{1}}:,
\end{aligned}
\ee
where the symbol $:~:$ means normal ordering.
Obviously
\be{cb}
\tilde c_{k_{2i}}\tilde b_{k_{2i-1}}=\cc\Delta^2 e^{iu(x_{k_{2i-1}}-x_{k_{2i}})}:\bigl(\psi^\dagger_1(k_{2i-1})\psi_1(k_{2i})
+\psi^\dagger_2(k_{2i-1})\psi_2(k_{2i})\bigr):.
\ee
Thus, we find
\begin{multline}\label{A-ell-2}
 A_\ell(u)=\cc^\ell\Delta^{2\ell}\sum_{N\ge k_{2\ell}>\dots>k_1\ge 1} \prod_{i=1}^\ell e^{iu(x_{k_{2i}}-x_{k_{2i-1}})}\\
 \times :\prod_{i=1}^{\ell-1}\bigl(\psi^\dagger_1(k_{2i})\psi_1(k_{2i+1})+\psi^\dagger_2(k_{2i})\psi_2(k_{2i+1})\bigr)
 \begin{pmatrix}
 \psi^\dagger_1(k_{2\ell})\psi_1(k_{1}) & \psi^\dagger_1(k_{2\ell})\psi_2(k_{1})\\
 \psi^\dagger_2(k_{2\ell})\psi_1(k_{1})& \psi^\dagger_2(k_{2\ell})\psi_2(k_{1})
  \end{pmatrix}:\;,
 \end{multline}
and
\begin{multline}\label{D-ell-2}
 D_\ell(u)=e^{iuL}\cc^\ell\Delta^{2\ell}\sum_{N\ge k_{2\ell}>\dots>k_1\ge 1} \prod_{i=1}^\ell e^{-iu(x_{k_{2i}}-x_{k_{2i-1}})}\\
 \times :\prod_{i=1}^{\ell}\bigl(\psi^\dagger_1(k_{2i-1})\psi_1(k_{2i})+\psi^\dagger_2(k_{2i-1})\psi_2(k_{2i})\bigr)
 :\;.
 \end{multline}

It remains to replace the sums over $k_i$ by integrals via \eqref{sum-int}. It is convenient to set $x_{k_{2i}}=z_i$ and $x_{k_{2i-1}}=y_i$. Then
\begin{multline}\label{A-ell-cont}
 A_\ell(u)=\cc^\ell\int_0^L \prod_{i=1}^\ell \left\{ e^{iu(z_{i}-y_{i})}\,dz_i\,dy_i\right\}\Theta_\ell(\bar z,\bar y)
 \\
 \times :\prod_{i=1}^{\ell-1}\bigl(\Psi^\dagger_1(z_{i})\Psi_1(y_{i+1})+\Psi^\dagger_2(z_{i})\Psi_2(y_{i+1})\bigr)
 \begin{pmatrix}
 \Psi^\dagger_1(z_{\ell})\Psi_1(y_{1}) & \Psi^\dagger_1(z_{\ell})\Psi_2(y_{1})\\
 \Psi^\dagger_2(z_{\ell})\Psi_1(y_{1})& \Psi^\dagger_2(z_{\ell})\Psi_2(y_{1})
  \end{pmatrix}:\;,
 \end{multline}
and
\begin{equation}\label{D-ell-cont}
 D_\ell(u)=e^{iuL}\cc^\ell\int_0^L \prod_{i=1}^\ell \left\{ e^{-iu(z_{i}-y_{i})}\,dz_i\,dy_i\right\}\Theta_\ell(\bar z,\bar y)
 :\prod_{i=1}^{\ell}\bigl(\Psi^\dagger_1(y_{i})\Psi_1(z_{i})+\Psi^\dagger_2(y_{i})\Psi_2(z_{i})\bigr):\;,
 \end{equation}
where
\be{Theta}
\Theta_\ell(\bar z,\bar y)=\theta(z_\ell-y_\ell)\prod_{i=1}^{\ell-1}\theta(y_{i+1}-z_i) \theta(z_i-y_i).
\ee

The anti-diagonal blocks of the monodromy matrix can be found exactly in the same manner. Setting
$n=2\ell+1$ in \eqref{Tnu-1} we find
\be{Tell-adiag}
\cc^{\ell+1/2} T_{2\ell+1}(u)=\begin{pmatrix}0& B_\ell(u) \\ C_\ell(u)&0 \end{pmatrix},
\ee
where
\begin{multline}\label{C-ell-cont}
 C_\ell(u)=ie^{iuL}\cc^{\ell+1/2}\int_0^L \prod_{i=1}^\ell \left\{ e^{iu(z_{i}-y_{i})}\,dz_i\,dy_i\right\}e^{-iuy_{\ell+1}}\,dy_{\ell+1}
 \theta(y_{\ell+1}-z_\ell)\Theta_\ell(\bar z,\bar y)\\
 \times
 :\prod_{i=1}^{\ell}\bigl(\Psi^\dagger_1(z_{i})\Psi_1(y_{i+1})+\Psi^\dagger_2(z_{i})\Psi_2(y_{i+1})\bigr)\cdot
 \begin{pmatrix}\Psi_{1}(y_1);& \Psi_{2}(y_1)
\end{pmatrix}:\;,
 \end{multline}
and
\begin{multline}\label{B-ell-cont}
 B_\ell(u)=-i\cc^{\ell+1/2}\int_0^L \prod_{i=1}^\ell \left\{ e^{-iu(z_{i}-y_{i})}\,dz_i\,dy_i\right\}e^{iuy_{\ell+1}}\,dy_{\ell+1}
 \theta(y_{\ell+1}-z_\ell)\Theta_\ell(\bar z,\bar y)\\
 \times
 :\prod_{i=1}^{\ell}\bigl(\Psi^\dagger_1(y_{i})\Psi_1(z_{i})+\Psi^\dagger_2(y_{i})\Psi_2(z_{i})\bigr)\cdot
  \begin{pmatrix}\Psi_{1}^\dagger(y_{\ell+1})\\ \Psi_{2}^\dagger(y_{\ell+1})
\end{pmatrix}:\;.
 \end{multline}

Thus, we have obtained the explicit series representation for the monodromy matrix entries $T_{ij}(u)$ in terms of the local Bose fields.
This series is formal, and we do not study the problem of its convergence. It is easy to see, however, that if we introduce a vector
\be{Phi}
|\Phi_{a,b}\rangle=\int_0^L \,dx_1,\dots,\,dx_a\,dy_1,\dots,\,dy_b \; \Phi_{a,b}(x_1,\dots,x_a;y_1,\dots,y_b)\prod_{i=1}^a\Psi^\dagger_1(x_i)\prod_{j=1}^b\Psi^\dagger_2(y_j)|0\rangle,
\ee
where $\Phi_{a,b}(x_1,\dots,x_a;y_1,\dots,y_b)$ is a continuous function within the integration domain, then the action
of any $T_{ij}(u)$  on $|\Phi_{a,b}\rangle$ turns into a finite sum.

\section{Mapping of fields\label{S-MF}}

Due to the invariance of the
$R$-matrix under transposition with respect to both spaces, the mapping
\be{def-psi}
\phi\bigl(T_{jk}(u)\bigr) = T_{kj}(u)
\ee
defines an antimorphism of the algebra \eqref{RTT} (see \cite{PakRS14b}). This mapping is a very convenient tool in studying form factors, because
it allows one to relate form factors of different operators. In the case of the TCBG model antimorphism \eqref{def-psi} agrees
with the following mapping of the Bose fields:
\be{psi-Psi}
\phi\bigl(\Psi_i(x)\bigr)=-\Psi^\dagger_i(L-x), \qquad \phi\bigl(\Psi^\dagger_i(x)\bigr)=-\Psi_i(L-x).
\ee
Indeed, consider, for example, how the mapping \eqref{psi-Psi} acts on the matrix elements $T_{jk}(u)$ for $j,k=1,2$.
Due to equations \eqref{Tell-diag}, \eqref{A-ell-cont} we have
\be{series-ij}
T_{jk}(u)=\sum_{\ell=0}^\infty \cc^\ell \bigl(T_{2\ell}\bigr)_{jk}(u), \qquad j,k=1,2,
\ee
where
\begin{equation}\label{Tjk-ell-cont}
\bigl(T_{2\ell}\bigr)_{jk}(u)=\int_0^L \prod_{n=1}^\ell \left\{ e^{iu(z_{n}-y_{n})}\,dz_n\,dy_n\right\}\Theta_\ell(\bar z,\bar y)
:\Psi^\dagger_j(z_{\ell})\Psi_k(y_{1})
  \prod_{n=1}^{\ell-1}\Bigl(\sum_{s=1}^2\Psi^\dagger_s(z_{n})\Psi_s(y_{n+1})\Bigr):\;.
 \end{equation}
Recall that due to the factor $\Theta_\ell(\bar z,\bar y)$  the integral in \eqref{Tjk-ell-cont} is taken over domain
$z_\ell>y_\ell>z_{\ell-1}>\dots>z_1>y_1$. Therefore all the operators in \eqref{Tjk-ell-cont} commute
with each other, and actually the normal ordering is not necessary. Acting on \eqref{Tjk-ell-cont}
with $\phi$ as in \eqref{psi-Psi} we obtain
\begin{multline}\label{psi-Tij-1}
\phi\bigl(\bigl(T_{2\ell}\bigr)_{jk}(u)\bigr)=\int_0^L \prod_{n=1}^\ell \left\{ e^{iu(z_{n}-y_{n})}\,dz_n\,dy_n\right\}\Theta_\ell(\bar z,\bar y)
:\Psi^\dagger_k(L-y_{1})\Psi_j(L-z_{\ell})\\
\times   \prod_{n=1}^{\ell-1}\Bigl(\sum_{s=1}^2\Psi^\dagger_s(L-y_{n+1})\Psi_s(L-z_{n})\Bigr):\;.
 \end{multline}
Now it is enough to change the integration variables
$z_n\to L-y_{\ell+1-n}$ and $y_n\to L-z_{\ell+1-n}$. Then we have
\begin{multline}\label{Theta-rep}
\Theta_\ell(\bar z,\bar y)\Bigr|_{\substack{z_n\to L-y_{\ell+1-n} \\y_n\to L-z_{\ell+1-n} }}
=\prod_{n=1}^{\ell-1}\theta(y_{\ell-n+1}-z_{\ell-n}) \prod_{n=1}^{\ell}\theta(z_{\ell-n+1}-y_{\ell-n+1})\\
=\prod_{i=1}^{\ell-1}\theta(y_{i+1}-z_{i}) \prod_{i=1}^{\ell}\theta(z_{i}-y_{i})=\Theta_\ell(\bar z,\bar y).
\end{multline}
It is also easy to see that
\begin{multline}\label{Prod-rep}
\prod_{n=1}^{\ell-1}\Bigl(\sum_{s=1}^2\Psi^\dagger_s(L-y_{n+1})\Psi_s(L-z_{n})\Bigr)
\Bigr|_{\substack{z_n\to L-y_{\ell+1-n} \\y_n\to L-z_{\ell+1-n} }}=
\prod_{n=1}^{\ell-1}\Bigl(\sum_{s=1}^2\Psi^\dagger_s(z_{\ell-n})\Psi_s(y_{\ell-n+1})\Bigr)\\
=\prod_{n=1}^{\ell-1}\Bigl(\sum_{s=1}^2\Psi^\dagger_s(z_{n})\Psi_s(y_{n+1})\Bigr).
\end{multline}
Thus, we arrive at
\begin{multline}\label{psi-Tij-2}
\phi\bigl(\bigl(T_{2\ell}\bigr)_{jk}(u)\bigr)=\int_0^L \prod_{n=1}^\ell \left\{ e^{iu(z_{n}-y_{n})}\,dz_n\,dy_n\right\}\Theta_\ell(\bar z,\bar y)
:\Psi^\dagger_k(y_{1})\Psi_j(z_{\ell})\\
\times   \prod_{n=1}^{\ell-1}\Bigl(\sum_{s=1}^2\Psi^\dagger_s(z_{n})\Psi_s(y_{n+1})\Bigr): = \bigl(T_{2\ell}\bigr)_{kj}(u) \;.
 \end{multline}
Similarly, using the explicit representations for other operators $T_{jk}(u)$ one can prove that \eqref{psi-Psi} implies \eqref{def-psi}.

\section{Zero modes\label{S-ZM}}

A method of calculating form factors of local operators in $GL(3)$-invariant models was developed  in \cite{PakRS15c}. This method
is based on the use of partial zero modes of the monodromy matrix entries $T_{ij}(u)$ in the composite model consisting of two partial
monodromy matrices \eqref{T-TT}.
In spite of this approach can be applied to a wide class of integrable models, it should be slightly modified in the case of the TCBG model. The matter is that it was  assumed in \cite{PakRS15c}  that the monodromy matrix $T(u)$ goes to the identity operator at $|u|\to\infty$.
This restriction is not very important, however it leads to minor changes in the case of the TCBG model.

Observe that a monodromy matrix $T^{(a)}(u)$ constructed by the $L$-operator \eqref{L-kulresh} possesses the property mentioned above. Indeed,
we can define local $L$-operators $L^{(a)}_n(u)$, ($n=1,\dots,N$) by equations \eqref{L-kulresh} and \eqref{p-kulresh}, where the operators
$a_k$ and $a^\dagger_k$ are respectively replaced with $a_k(n)$ and $a^\dagger_k(n)$ with the commutation relations
$[a_i(n),a^\dagger_k(m)]=\delta_{nm}\delta_{ik}$. Then we can set
\be{Ta-La}
T^{(a)}(u)=u^{-N}L^{(a)}_N(u)\dots L^{(a)}_1(u),
\ee
and this matrix obviously has an asymptotic expansion
\be{Ta-asy}
T^{(a)}(u)=\mathbf{I}+\frac cu T^{(a)}[0]+O(u^{-2}),\qquad u\to\infty.
\ee
Therefore we can define zero modes of this monodromy matrix in a standard way
\be{Ta-zero}
T^{(a)}[0]=\lim_{u\to\infty}\frac uc \bigl(T^{(a)}(u)-\mathbf{I}\bigr).
\ee

However, passing from  $L$-operator \eqref{L-kulresh} to $L$-operator \eqref{L-op} we have multiplied $L^{(a)}(u)$
by the matrix $J=\diag(1,1,-1)$ (see \eqref{IK-RK}). This led to the fact that the monodromy matrix $T(u)$ \eqref{T-L}
in the continuous limit has essential singularity at infinity. Therefore, in the case of the TCBG model, the definition of zero
modes needs to be clarified. We do it in this section and consider an asymptotic expansion of the monodromy matrix entries $T_{ij}(u)$ at large value of the argument. For this purpose we use the integral representations for $T_{ij}(u)$ obtained in section~\ref{S-RMM-BF}.

If $u\to\infty$, then the expansion for the monodromy matrix contains multiple integrals of quickly oscillating exponents.
Methods of calculating quickly oscillating integrals are well known (see e.g. \cite{Bru61,Erd56}). In our case
the integration domain of every integration variable is a finite interval $[0,L]$, therefore one of the simplest ways to obtain the asymptotic expansion of
$T_{ij}(u)$ is the integration by parts. Using this method one can easily show that single and double integrals  give $1/u$-behavior, while all
the terms with $\ell>1$ give contributions of order $o(u^{-1})$. Therefore, in order to find zero modes it is enough to take
only the first nontrivial terms of the expansion for $T(u)$. Then we have
\be{Tij}
T_{ij}(u)= \delta_{ij}+\cc\int_0^L e^{iu(z-y)}\theta(z-y)\Psi_i^\dagger(z)\Psi_j(y)\,dz\,dy+O(\cc^2),\qquad i,j=1,2,
\ee
\begin{equation}\label{T33}
 T_{33}(u)=e^{iuL}+\cc\,e^{iuL}\int_0^L e^{iu(y-z)}\theta(z-y)
\bigl(\Psi^\dagger_1(y)\Psi_1(z)+\Psi^\dagger_2(y)\Psi_2(z)\bigr)\,dz\,dy+O(\cc^2).
 \end{equation}
\begin{equation}\label{Ti3}
T_{i3}(u)=-i\sqrt{\cc}\int_0^L e^{iuy}\Psi_{i}^\dagger(y)\,dy  +O(\cc^{3/2}),\qquad i=1,2,
 \end{equation}
\begin{equation}\label{T3j}
T_{3j}(u)=i\sqrt{\cc}e^{iuL}\int_0^L e^{-iuy}\Psi_{j}(y)\,dy  +O(\cc^{3/2}),\qquad j=1,2,
 \end{equation}
All the terms denoted by $O(\cc^2)$ or $O(\cc^{3/2})$ give contributions $O(u^{-2})$ as $u\to\infty$, and therefore they are not important.
Integrating by parts we obtain
\be{Tij-1}
T_{ij}(u)= \delta_{ij}+\frac{i\cc}{u}\int_0^L \Psi_i^\dagger(y)\Psi_j(y)\,dy+O(u^{-2}),\qquad i,j=1,2,
\ee
\begin{equation}\label{T33-1}
 T_{33}(u)=e^{iuL}-\frac{i\cc}{u}\,e^{iuL}\int_0^L
\bigl(\Psi^\dagger_1(y)\Psi_1(y)+\Psi^\dagger_2(y)\Psi_2(y)\bigr)\,dy+O(u^{-2}).
 \end{equation}
\begin{equation}\label{Ti3-1}
T_{i3}(u)=-\frac{\sqrt{\cc}}{u}\left(e^{iuL}\Psi_{i}^\dagger(L)-\Psi_{i}^\dagger(0)\right)+O(u^{-2}),\qquad i=1,2,
 \end{equation}
\begin{equation}\label{T3j-1}
T_{3j}(u)=-\frac{\sqrt{\cc}}{u}\left(\Psi_{j}(L)-e^{iuL}\Psi_{j}(0)\right)+O(u^{-2}),\qquad j=1,2,
 \end{equation}

Now we define zero modes as follows:
\be{0Tij}
T_{ij}[0]=\lim_{u\to\infty}\frac uc(T_{ij}(u)-\delta_{ij})=-\int_0^L \Psi_i^\dagger(y)\Psi_j(y)\,dy, \qquad i,j=1,2,
\ee
(recall that $c=-i\cc$). This is the same definition as for the models considered in \cite{PakRS15c}. The zero mode
$T_{33}[0]$ is slightly differently:
\be{0T33}
T_{33}[0]=\lim_{u\to\infty}\frac uc\left(e^{-iuL} T_{33}(u)-1\right)=\int_0^L
\bigl(\Psi^\dagger_1(y)\Psi_1(y)+\Psi^\dagger_2(y)\Psi_2(y)\bigr)\,dy,
\ee
and thus, $T_{11}[0]+T_{22}[0]=-T_{33}[0]$.

Looking at \eqref{Ti3-1}, \eqref{T3j-1} we see that actually we have two types of zero modes for these operators. We call
them left and right zero modes and denote respectively by $T^{(\text{\tiny L})}_{ij}[0]$ and $T^{(\text{\tiny R})}_{ij}[0]$.
Then
\be{0Ti3}
\begin{aligned}
&T^{(\text{\tiny R})}_{k3}[0]=\lim_{u\to-i\infty}e^{-iuL}\frac uc \;T_{k3}(u)=\frac1{i\sqrt{\cc}}\;\Psi^\dagger_k(L),\\
&T^{(\text{\tiny L})}_{k3}[0]=\lim_{u\to+i\infty}\frac uc \;T_{k3}(u)=-\frac1{i\sqrt{\cc}}\;\Psi^\dagger_k(0),
\end{aligned}
 \qquad k=1,2,
\ee
and
\be{0T3j}
\begin{aligned}
&T^{(\text{\tiny R})}_{3j}[0]=\lim_{u\to+i\infty}\frac uc \;T_{3j}(u)=\frac1{i\sqrt{\cc}}\;\Psi_j(L), \\
&T^{(\text{\tiny L})}_{3j}[0]=\lim_{u\to-i\infty}e^{-iuL}\frac uc \;T_{3j}(u)=-\frac1{i\sqrt{\cc}}\;\Psi_j(0),
\end{aligned}
\qquad j=1,2.
\ee

The sums $T^{(\text{\tiny L})}_{ij}[0]+T^{(\text{\tiny R})}_{ij}[0]$ play the same role as the zero modes
of the monodromy matrix of the type \eqref{Ta-La}, \eqref{Ta-asy}. It is known, in particular \cite{MuhTV06,PakRS15a}, that
some of zero modes $T_{ij}[0]$ annihilate on-shell Bethe vectors:
\be{ZM-Ann}
T^{(a)}_{ij}[0]\mathbb{B}_{a,b}(\bla,\bmu)=0,\qquad i>j.
\ee
Similarly one can check that
\be{ZM-AnnS}
(T^{(\text{\tiny L})}_{3j}[0]+T^{(\text{\tiny R})}_{3j}[0])\mathbb{B}_{a,b}(\bla,\bmu)=0,\qquad
 j \ne 3,
\ee
provided $\mathbb{B}_{a,b}(\bla,\bmu)$ is an on-shell vector. In order to prove
\eqref{ZM-AnnS} it is sufficient to use the formulas of the action of $T_{ij}(u)$ onto Bethe vectors \cite{BelPRS12c} and to
consider there the limits $u\to\pm i\infty$ like in \eqref{0T3j}.

Finally, the obtained formulas for the zero modes allow us to study form factors of the local operators in the framework of the composite
model \eqref{T-TT}. Indeed, let in \eqref{T-TT} the partial monodromy matrix $T^{(1)}(u)$ corresponds to an interval $[0,x]$, where
$x$ is a fixed point of the interval $[0,L]$. Then the partial zero modes $T^{(1)}_{ij}[0]$
and $T^{(1;\text{\tiny R})}_{ij}[0]$ are given by \eqref{0Tij}--\eqref{0T3j}, where
one should replace everywhere $L$ by $x$. In particular, we obtain
\be{ZM-TCBG}
\begin{aligned}
&\Psi_i^\dagger(x)\Psi_j(x)=-\frac{d}{dx}T^{(1)}_{ij}[0]=\frac1{i\cc}\frac{d}{dx}\lim_{u\to\infty} u(T^{(1)}_{ij}(u)-\delta_{ij}), \qquad i,j=1,2,\\
 &\Psi_j(x)=i\sqrt{\varkappa}\;T^{(1;\text{\tiny R})}_{3j}[0]=\frac1{\sqrt{\cc}}\lim_{u\to+i\infty} u \;T^{(1)}_{3j}(u), \qquad j=1,2,\\
&\Psi^\dagger_k(x)=i\sqrt{\varkappa}\;T^{(1;\text{\tiny R})}_{k3}[0]=\frac1{\sqrt{\cc}}\lim_{u\to-i\infty}e^{-iux}  u \;T^{(1)}_{k3}(u), \qquad k=1,2.
\end{aligned}
\ee
Thus, the problem of calculating the form factors of the local operators in the TCBG model is reduced to the evaluating the
form factors of the partial zero modes $T^{(1)}_{ij}[0]$
and $T^{(1;\text{\tiny R})}_{ij}[0]$. 

\section*{Conclusion}

In this paper we gave a description of the TCBG model in the framework of the algebraic Bethe ansatz. The main goal was to prove
that the lattice $L$-operator \eqref{L-op} correctly describes the TCBG model in the continuous limit and allows one to find the zero modes
of the monodromy matrix $T(u)$. This goal is successfully achieved. In order to calculate the form factors of the fields $\Psi_i(x)$,
$\Psi_i^\dagger(x)$, and their combinations $\Psi_i^\dagger(x)\Psi_j(x)$ we can use now the method of \cite{PakRS15c}.
Actually, a part of results can be predicted already now. Indeed, the definition \eqref{0Tij} of the zero modes $T_{ij}[0]$ for $i,j=1,2$  coincides
with the definition used in \cite{PakRS15c}. Therefore the  form factors of the operators  $\Psi_i^\dagger(x)\Psi_j(x)$ in fact are already
computed. The calculation of the form factors of the fields $\Psi_i(x)$ and $\Psi_i^\dagger(x)$ should be slightly modified. However, in this case the modification affects only the limit $u\to\infty$, but does not affect determinant representations for the partial zero modes.
We consider this question in details in our further publication.

\section*{Acknowledgements}

It is a great pleasure for me to thank my colleagues S. Pakuliak and E. Ragoucy for numerous and fruitful discussions.
This work was supported by the RSF under a grant 14-50-00005.

\end{document}